\documentclass[12pt]{article}
\pdfoutput=1
\usepackage{epsf,color,colordvi}
\usepackage{graphicx}
\usepackage{amsmath}
\usepackage{verbatim}
\usepackage{amssymb}
\usepackage[most]{tcolorbox}
\usepackage{epsfig}
\usepackage{cite}
\usepackage{fontenc}
\usepackage{graphicx}
\usepackage{float}
\usepackage[lofdepth,lotdepth,caption=false]{subfig}
\usepackage{color}
\usepackage{booktabs}
\usepackage{tabularx}
\usepackage{multirow}
\usepackage{longtable}
\usepackage{lscape}
\usepackage[font=small,skip = 0pt]{caption}
\usepackage{dcolumn}
\usepackage{rotating}

\usepackage{subcaption,graphicx}
\usepackage[english]{babel}
\usepackage{soul,xcolor}
\usepackage[font=scriptsize]{caption}
\usepackage{braket}
\usepackage{bbold}
\usepackage{subfig}
\usepackage{braket}
\usepackage{titlesec}
\usepackage{placeins}
\usepackage{hyperref}
\usepackage{caption}
\usepackage{slashed}
\usepackage{epsf,color,colordvi}
\usepackage{amssymb}
\usepackage{fontenc}
\usepackage[lofdepth,lotdepth,caption=false]{subfig}
\usepackage{booktabs}
\usepackage{tabularx}
\usepackage{multirow}
\usepackage{longtable}
\usepackage{lscape}
\usepackage{dcolumn}
\usepackage{rotating}
\usepackage[autostyle]{csquotes}
\usepackage{xcolor}
\usepackage[normalem]{ulem}
\usepackage[export]{adjustbox}

\allowdisplaybreaks[4] 
\makeatletter
\renewcommand{\fnum@table}{\textbf{\tablename~\thetable}}
\renewcommand{\fnum@figure}{\textbf{\figurename~\thefigure}}
\makeatother

\newcounter{myenumi}

\renewcommand{\themyenumi}{\roman{myenumi}}
{\end{list}}

\newlength{\myem}
\newcounter{mysubequation}[equation]

\makeatletter
\renewcommand{\section}{\@startsection{section}{1}{0em}{-\baselineskip}%
{\baselineskip}{\normalfont\large\bfseries}}
\renewcommand{\subsection}%
{\@startsection{subsection}{2}{0em}{-0.7\baselineskip}%
{0.7\baselineskip}{\normalfont\bfseries}}
\makeatother

\newcommand{\bi}{\begin{itemize}}
\newcommand{\ei}{\end{itemize}}

\def\beq{\begin{equation}}
\def\eeq{\end{equation}}

\newcommand{\bea}{\begin{eqnarray}}
\newcommand{\eea}{\end{eqnarray}}

\newcommand{\ie}{{\it i.e.}}

\def\epsilon{\varepsilon}

\def\<{\langle}
\def\>{\rangle}

\def\dfrac#1#2{{\displaystyle\frac{#1}{#2}}}

\def\lsim{\mathrel{\rlap{\lower4pt\hbox{\hskip1pt$\sim$}}
    \raise1pt\hbox{$<$}}}         
\def\gsim{\mathrel{\rlap{\lower4pt\hbox{\hskip1pt$\sim$}}
    \raise1pt\hbox{$>$}}}         

\newcommand{\dacp}[1]{\ensuremath{\delta [\Delta P^{CP/T}_{\alpha\beta}]}}

\newcommand{\pbarab}[1]{\ensuremath{{ P}_{\bar{\alpha} \bar{\beta}} }}
\newcommand{\pbarba}[1]{\ensuremath{{ P}_{\bar{\beta} \bar{\alpha}} }}
\newcommand{\acpab}[1]{\ensuremath{A^{CP}_{\alpha \beta}}}
\newcommand{\acpaa}[1]{\ensuremath{A^{CP}_{\alpha \alpha}}}
\newcommand{\ataa}[1]{\ensuremath{A^{T}_{\alpha \alpha}}}
\newcommand{\acpba}[1]{\ensuremath{A^{CP}_{\beta \alpha}}}
\newcommand{\atab}[1]{\ensuremath{{A}^{T}_{\alpha \beta}}}
\newcommand{\atba}[1]{\ensuremath{{A}^{T}_{\beta \alpha}}}
\newcommand{\acptab}[1]{\ensuremath{A^{CPT}_{\alpha \beta}}}

\newcommand{\dpcpme}[1]{\ensuremath{\Delta { P}^{CP}_{\mu e} }}
\newcommand{\dpcpmt}[1]{\ensuremath{\Delta { P}^{CP}_{\mu \tau} }}
\newcommand{\dpcpet}[1]{\ensuremath{\Delta { P}^{CP}_{e \tau } }}
\newcommand{\dpcpee}[1]{\ensuremath{\Delta { P}^{CP}_{e e} }}
\newcommand{\dpcpmm}[1]{\ensuremath{\Delta { P}^{CP}_{\mu \mu} }}
\newcommand{\dpcptt}[1]{\ensuremath{\Delta { P}^{CP}_{\tau \tau } }}
\newcommand{\dptme}[1]{\ensuremath{\Delta { P}^{T}_{\mu e}}}
\newcommand{\dptmt}[1]{\ensuremath{\Delta { P}^{T}_{\mu \tau}}}
\newcommand{\dptet}[1]{\ensuremath{\Delta { P}^{T}_{e \tau}}}
\newcommand{\dpcptab}[1]{\ensuremath{\Delta { P}^{CPT}_{\alpha \beta} }}

\begin{document}
\begin{titlepage}

\vspace*{-3.cm}
\begin{flushright}

\end{flushright}


\renewcommand{\thefootnote}{\fnsymbol{footnote}}
\setcounter{footnote}{-1}

{\begin{center}
{\large\bf 
Distinguishing between Dirac and Majorana neutrinos using temporal correlations
\\[0.2cm]
}
\end{center}}

\renewcommand{\thefootnote}{\alph{footnote}}

\vspace*{.8cm}
\vspace*{.3cm}
{
\begin{center} 

   {\sf           Bhavya Soni$^{\S}$\,  \footnote[1]{\makebox[1.cm]{Email:} soni.16@iitj.ac.in},
                }
        {\sf           Sheeba Shafaq$^{\P}$\,  \footnote[2]{\makebox[1.cm]{Email:} sheebakhawaja7@gmail.com }
        }
and                                           
            {\sf            Poonam Mehta$^{\Upsilon}$\,\footnote[3]{\makebox[1.cm]{Email:} pm@jnu.ac.in}
}
\end{center}
}
\vspace*{0cm}
{\it 
\begin{center}
$^\S$\, Indian Institute of Technology, Jodhpur, Jodhpur 342037, India  \\
$^\P$\, 508-Rose Enclave, Shivpora-B, Srinagar, Jammu and Kashmir 190004, India  \\
$^\Upsilon$\, School of Physical Sciences, Jawaharlal Nehru University, 
      New Delhi 110067, India  \\

\end{center}
}

{\Large 
\bf
 \begin{center} Abstract  
  
\end{center} 
 }
In the context of two flavour neutrino oscillations, it is  understood that the $2\times 2$ mixing matrix is parameterized by one angle and  a Majorana phase. However, this phase does not impact the oscillation probabilities in vacuum or in matter with constant density. Interestingly, the  Majorana phase becomes relevant when we describe neutrino oscillations along with neutrino decay. This is due to the fact that effective Hamiltonian has Hermitian and anti-Hermitian components which cannot be simultaneously diagonalized (resulting in decay eigenstates being different from the mass eigenstates). We consider the $\cal PT$ symmetric non-Hermitian Hamiltonian describing two flavour neutrino case and study the violation of Leggett-Garg Inequalities (LGI) in this context for the first time. We demonstrate that temporal correlations in the form of LGI allow us to probe whether neutrinos are Dirac or Majorana. We elucidate the role played by the mixing and decay parameters on the extent of violation of LGI. We emphasize that for optimized choice of parameters, the difference in $K_4$ ($K_3$) for Dirac and Majorana case is  $\sim 15\%$ ($\sim 10\%$). 

\vspace*{.5cm}

\end{titlepage}

\newpage

\renewcommand{\thefootnote}{\arabic{footnote}}
\setcounter{footnote}{0}

\section{Introduction} \label{sec:1}

In a classic paper,  Bender and Boettcher~\cite{Bender:1998ke} (see also \cite{Bender:2007nj}) invoked a very insightful idea of $\cal PT$ symmetry in non-Hermitian Hamiltonians and suggested that $\cal PT$ symmetry led to the real and positive spectra, thereby replacing the condition of self-adjointness to ensure real eigenvalues. Their work has triggered a lot of work  across diverse fields spanning optics to nuclear and particle physics. Using the simplest example of a  two-level quantum system, the intricacies of the $\cal PT$ symmetric non-Hermitian Hamiltonians have been extensively studied in different contexts~\cite{Bender:2002yp,Berry:2004} leading to useful insights.

It is well-known that the two flavor neutrino system is equivalent to a two state quantum system in the ultra-relativistic limit (for equal and fixed momenta of two neutrinos)~\cite{Mehta:2009ea}.
The consequences of $\cal PT$ symmetric non-Hermitian Hamiltonian in the context of two flavour neutrino oscillations have been widely investigated in the recent times. 
Ohlsson~\cite{Ohlsson:2015xsa} developed an approach to extend the ordinary two flavour neutrino oscillation formalism in matter for the case  non-Hermitian $\cal PT$ symmetric effective Hamiltonian. 
Ohlsson and Zhou~\cite{Ohlsson:2019noy}  calculated the transition probabilities for flavour eigenstates for two flavour neutrinos and discussed some implications of the $\cal PT$ broken phase.  Later the authors~\cite{Ohlsson:2020gxx} developed  the density matrix formalism for $\cal PT$ symmetric non-Hermitian open quantum systems in the presence of Lindblad decoherence. Chattopadhyay et al.~\cite{Chattopadhyay:2021eba} showed that the Hermitian and anti-Hermitian components of the effective two flavour Hamiltonian cannot be simultaneously diagonalized thereby resulting in decay eigenstates being different from the mass eigenstates. 
For two flavour neutrino oscillations, it is generally understood that the $2\times 2$ mixing matrix is parameterized by one angle and a Majorana phase however this  phase does not impact the oscillation probabilities in vacuum or in matter with constant density~\cite{Giunti:2010ec,Mehta:2009ea,Mehta:2009xm}. 
Following~\cite{Chattopadhyay:2021eba}, Dixit  et al.~\cite{Dixit:2022izn} showed that the Majorana phase in the mixing matrix can appear at the level of detection  probabilities if proper treatment is carried out for the case of neutrino oscillations along with decay. 
Naumov et al.~\cite{Naumov:2020tba} considered  non-Hermitian Hamiltonian for three neutrino case and obtained a relation between the neutrino oscillation parameters in vacuum and their counterparts in matter. The analytic treatment of neutrino oscillations and decay was carried out for the three flavour case in~\cite{Chattopadhyay:2022ftv}.

Quantum mechanics has been extremely successful however some concerns are raised while discussing the applicability of quantum mechanics to the macroscopic world. In {1935}, Einstein, Podolsky, and Rosen (EPR)~\cite{Einstein:1935rr} questioned if the quantum mechanical description of the world is indeed complete. Later, in 1964, using the idea of local realism and spatially separated systems, Bell~\cite{Bell:1964fg} introduced the famous Bell's inequalities which could allow for a distinction between classical and quantum correlations for such a system.  Violation of Bell's inequality has been experimentally tested in different branches of physics. In turn, these tests allow us to examine the compatibility of local hidden variable theories with quantum mechanics. In a profound  development, Leggett and Garg~\cite{Leggett:1985zz} (see \cite{Nori:2014lgi} for a review) introduced the Leggett-Garg inequalities (LGI), which involve performing measurements on a single system at different points in time. 
This approach offers a distinct possibility to test the applicability of quantum mechanics as we go from microscopic to macroscopic world. 
Violation of LGI for non-Hermitian $\cal PT$-symmetric dynamics (via a sequence of dichotomic projective measurements which are carried out at different time intervals)  has been studied in~\cite{Karthik:2019wbp,varma2021temporal}.

Unlike their photonic counterparts, neutrinos exhibit quantum coherence over astronomical length scales. This makes neutrinos unique probes of foundational issues related to quantum mechanics and in particular, LGI. Violation of LGI was studied in the context of oscillations of neutral kaons and neutrinos~\cite{Gangopadhyay:2013aha,Gangopadhyay:2017nsn}. The three flavour analysis was carried out assuming the stationarity condition~\cite{Naikoo:2017fos} and relaxing it~\cite{Naikoo:2019eec}. Certain other forms of LGI in subatomic systems have been studied in~\cite{Naikoo:2019gme}. In fact, two neutrino experiments have demonstrated violation of LGI in their data at high level of significance. The  Main Injector Neutrino Oscillation Search (MINOS) experiment was the first to report $\sim 6\sigma$ violation of LGI over a  macroscopic length scale of $735$ km~\cite{Formaggio:2016cuh}. Even though Bell's tests (or its temporal analogue) have been performed in different contexts, the MINOS experiment provided the longest ever distance over which such a test had been carried out. In the Daya Bay reactor experiment, $\sim 6\sigma$ violation of LGI was reported in the data~\cite{Fu:2017hky}. 
Note that these tests were performed assuming two neutrino states only.

Neutrino oscillations in two and three flavours can be described in terms of qubits and qutrits used in quantum information theory~\cite{Mehta:2009ea,Mehta:2009xm,Jha:2022yik} and  entanglement in neutrino oscillations has been studied in~\cite{KumarJha:2020pke,Siwach:2022xhx,Ettefaghi:2023zsh}.  Quantum studies of neutrinos have been implemented on IBMQ processors~\cite{Jha:2021itm}. No signaling in time has also been studied in the context of neutrino oscillations~\cite{Blasone:2022iwf}. 
There are some other measures to examine coherence in neutrino oscillations such as contextuality~\cite{Richter-Laskowska:2018ikv}, $l_1$ norm of coherence~\cite{Dixit:2019swl},  entropic uncertainty relations~\cite{Wang:2020vdm}, quantum spread complexity~\cite{Dixit:2023fke} and quantum mismatch~\cite{Chattopadhyay:2023xwr}. 
Tools of quantum resource theory~\cite{Song:2018bma, Ming:2020nyc}
have also been used to quantify the quantumness of  neutrino oscillations. %
Impact of new physics such as non-standard interactions~\cite{Shafaq:2020sqo,Dixit:2019swl,Sarkar:2020vob,Yadav:2022grk} and damping effects~\cite{Shafaq:2021lju,Richter:2017toa} on quantum correlations in neutrino oscillations have also been studied.

That neutrinos, being electrically neutral, could possibly be of Majorana type was proposed in 1937~\cite{Majorana2020} (see~\cite{Pal:2010ih,Bilenky:2020wjn} for a review). Majorana's  insightful idea has triggered extensive theoretical and experimental work. The smoking gun signal could come from the so-called neutrinoless double $\beta$-decay process which violates lepton number by two units and is proportional to the Majorana mass of the neutrino. Several laboratories around the world host experiments to detect neutrinoless double $\beta$-decay, with no success so far~\cite{Dolinski:2019nrj}.

In the present work, we investigate the violation of LGI in the context of non-Hermitian $\cal PT$ symmetric two flavour neutrino system. We  demonstrate that we can probe the nature of neutrinos (\ie, Dirac or Majorana character) via the extent of violation of LGI.
As far as leptonic mixing is concerned,  there is only one Majorana phase~\cite{Giunti:2010ec,Pal:2010ih} in the two flavour scenario. For three (or more) flavours, we can have one (or more) Dirac-type phases and two (or more) Majorana phases~\footnote{{For $N$ generations of leptons, there are $(N-1)(N-2)/2$ Dirac phases and $(N-1)$  Majorana phases.}}.  
We elucidate the role played by the mixing and decay  parameters on the extent of violation of LGI. For certain favourable choice of parameters and examining the dependence on the Majorana phase, we show that we can discriminate between Dirac and Majorana case.

This article is organized as follows. In Sec.~\ref{sec:2}, we describe  the framework of $\cal PT$-symmetric non-Hermitian Hamiltonian in the context of two flavour neutrino oscillations. In Sec.~\ref{sec:3}, we present our results on LGI violation. Finally, we conclude in Sec.~\ref{Sec:4}.


\section{Framework}\label{sec:2}

\subsection{Two flavour neutrino oscillations with Hermitian Hamiltonian}\label{2.1}

A general $2\times2$ unitary mixing matrix can be expressed as~\cite{Giunti:2010ec}
\begin{eqnarray}
    U
    =
    \begin{pmatrix}
    \cos\theta e^{i\omega_{1}} & \sin\theta e^{i(\omega_{1}+\phi)}
    \\
    - \sin\theta e^{i(\omega_{2}-\phi)} & \cos\theta e^{i\omega_{2}}
    \end{pmatrix}
    \,.
    \label{U}
\end{eqnarray}
We note that $U$ is parameterized by one angle and three phases.  It is possible to rephase the two Dirac charged-lepton fields (without affecting the kinetic and mass Lagrangians as well as Lagrangians of other interactions to which charged leptons take part) and eliminate two of these phases. However, it is not possible to rephase the Majorana field as Majorana mass term is not invariant under rephasing of the field. Thus, one of the phases remains physical and is  referred to  as the ``Majorana phase"~\cite{Majorana2020} (see also~  \cite{Bilenky:1980cx,Doi:1980yb,Schechter:1980gk}).

\renewcommand{\labelenumi}{(\theenumi)}
\renewcommand{\theenumi}{\arabic{enumi}}

Now to address the question of observability of the Majorana phase, let us rephase the charged-lepton fields as~\footnote{This choice is not unique and we refer the reader to \cite{Giunti:2010ec} for details.}
 $ e_{L}(x) \to e^{i\omega_{1}} e_{L}(x)
   $ and $ 
    \mu_{L}(x) \to e^{i(\omega_{2}-\phi)} \mu_{L}(x)
    $.

This leads to the following form of the mixing matrix 
\begin{eqnarray}
    U
    =
    \begin{pmatrix}
    \cos\theta & \sin\theta e^{i\phi}
    \\
    - \sin\theta & \cos\theta e^{i\phi}
    \end{pmatrix}
    =
    \begin{pmatrix}
    \cos\theta & \sin\theta
    \\
    - \sin\theta & \cos\theta
    \end{pmatrix}
    \begin{pmatrix}
    1 & 0
    \\
    0 & e^{i\phi}
    \end{pmatrix}
    =
    R(\theta) D(\phi)
    \,,
    \label{U3}
\end{eqnarray}
where the Majorana phase $\phi$ has been factorized as a diagonal matrix $ D(\phi) = \text{diag}(1,e^{i\phi}) $
on the right side of the mixing matrix. 
While there were some claims that Majorana phase may be observable in neutrino oscillation experiments (with an initial beam described by superposition of flavors~\cite{Adhikari:2009ja}), it is clear that the Majorana phase can not appear at the level of oscillation probabilities in the context of two flavour neutrino oscillations~\cite{Giunti:2010ec}. 

We can  qualify this statement further for a generalized situation. It is known that neutrino mixing and neutrino decay can be described by non-Hermitian quantum dynamics. In this scenario, it is possible to visualize the effects of the Majorana phase at the level of detection probabilities. The main reason is as follows. The mass eigenstates and decay eigenstates are not the same~\cite{Chattopadhyay:2021eba} and therefore if the decay term in the Hamiltonian has off-diagonal entries, we can get a unique opportunity to see the effect of  Majorana phase through detection probabilities of neutrinos~\cite{Dixit:2022izn}.

\subsection{Two flavour neutrino oscillations with decay and PT symmetric non-Hermitian Hamiltonians}
\label{2.2}

A general non-Hermitian Hamiltonian $\mathcal{H}$ can be expressed as $\mathcal{H} = \mathcal{H}_+ + \mathcal{H}_- $ with $\mathcal{H}_\pm = (\mathcal{H} \pm \mathcal{H}^\dagger)/2$ with $\mathcal{H}_+$ being Hermitian and $\mathcal{H}_-$ being anti-Hermitian, respectively. For decay, $\mathcal{H}$ is usually written on the Weisskopf-Wigner form~\cite{Bertlmann:2006fn},
\begin{eqnarray}
    \mathcal{H} = \mathcal{M} - i \Gamma/2,
    \label{H}
\end{eqnarray}
where the Hermitian matrices $\mathcal{M}$ and $\Gamma/2$ have the form
	  \begin{eqnarray}
	      \mathcal{M} = 
	  	\begin{pmatrix}
	  		a_1                &0\\
	  		0                & a_2
	  	\end{pmatrix},~~~ 
  	    \Gamma/2 = 
  	    \begin{pmatrix}
  		        b_1                 &\frac{1}{2}\eta e^{i \xi}\\
  		        \frac{1}{2}\eta e^{-i \xi}                & b_2
  	    \end{pmatrix}\,,
       \label{H1}
	  \end{eqnarray}
   where $a_i, b_i, \eta$ and $ \xi$ are real with $a_2 - a_1 = \Delta m^2/2E$. Note that $\Delta m^2 = m_2^2-m_1^2$ denotes the  mass-squared difference between the two states and $E$ is the energy of the neutrinos. Since $\Gamma$ is positive semidefinite, it follows that $b_i \geq 0$ and $\eta^2 \leq 4 b_1 b_2$. 
   We consider $b_1=b_2=b$ to make second matrix $\cal PT$ symmetric. We assume $\eta \ll |a_2-a_1|$ for sake of simplicity.
If $\Gamma$ is diagonal ({\it i.e.}, $\eta = 0$), the decay eigenbasis is the same as the mass eigenbasis and the Majorana phase $\phi$ disappears from neutrino evolution equations. But, if $\Gamma$ is non-diagonal ({\it i.e.}, $\eta \neq 0$) the mass eigenstates are not the same as decay eigenstates ({\it i.e.}, $\sigma_z$ and $\Gamma$ do not  commute). As a consequence, Majorana phase appears at the level of oscillation probabilities as shown in~\cite{Dixit:2022izn}. We would like to remark that a complex non-Hermitian Hamiltonian can also be realised when absorption effects play a role giving rise to complex indices of refraction~\cite{Naumov:2001ci}.

The Hamiltonian can be expressed as~\cite{Dixit:2022izn} 
\begin{eqnarray}
   {\mathcal H} &=& 
   \left[\dfrac{(a_1+a_2)}{2} \sigma_0 - \dfrac{(a_2-a_1)}{2} \sigma_z - \dfrac{i}{2}\left((b_1 + b_2)\sigma_0+ \vec{\sigma}.\vec{\Gamma}\right)\right] \,,
    \label{H3}
\end{eqnarray}
where $\vec{\Gamma} = [\eta \cos \xi, -\eta \sin \xi, -(b_2-b_1)]$. This clearly shows that $[{\mathcal H}, \Gamma] \neq 0$. 

The oscillation probabilities for the case of neutrino oscillation and decay are 
\begin{eqnarray}
    P_{e\mu} &=& e^{-2bt} \left[P_{e\mu}^{\rm vac} + 2\eta \sin(\xi-\phi) \mathcal{B}\right]\, ,\nonumber \\
    P_{\mu e} &=& e^{-2bt} \left[P_{\mu e}^{\rm vac} - 2\eta \sin(\xi-\phi) \mathcal{B}\right]\, ,\nonumber\\
    P_{ee}&=& e^{-2bt} \left[ P_{ee}^{\rm vac} - \eta \cos(\xi-\phi) \mathcal{A}\right] \,, \nonumber \\
    P_{\mu\mu} &=&  e^{-2bt} \left[P_{\mu\mu}^{\rm vac} + \eta \cos(\xi-\phi) \mathcal{A}\right]\, .  
    \label{P}
    \end{eqnarray}
where,  $\mathcal{A}$  are $\mathcal{B}$ given by 
\begin{eqnarray}
\label{AB}
   	\mathcal{A} &=& \dfrac{\sin (2 \theta ) \sin \left[\left(a_2-a_1\right)t\right]}{(a_2-a_1)} \,,\nonumber\\ 
   	 \mathcal{B} &=& \dfrac{\sin (2 \theta ) \sin ^2\left[\dfrac{1}{2}  (a_2-a_1) t \right]}{(a_2-a_1)} \,.
   \end{eqnarray}
   
\begin{figure}[ht!]
\includegraphics[width=1\textwidth]{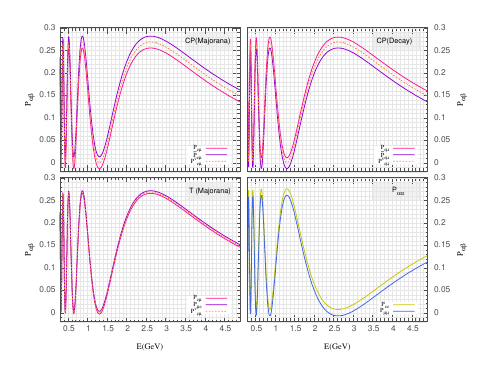} 
\caption{\label{fig:violations} 
Oscillation probability (Eq.~\ref{P}) is plotted as a function of energy for a fixed baseline of $L = 1300$ km. The different panels correspond to 
(a) $\cal CP$ violation due to $\phi$ (top left), (b) $\cal CP$ violation due to decay, $\xi$ (top right), (c) $\cal T$ violation (bottom left) and (d) survival probabilities for the different flavours (bottom right). %
Note that $\Delta m^2 = 2.5\times 10^{-3}$  eV$^2$, $\theta = 45^{\circ}$, $b=10^{-4}$ eV$^2$, $\eta\ = 10^{-4} $eV$^2$, $\xi\ = \pi/5$, $\phi\ = \pi/4$.}
\end{figure}

   The vacuum probability expressions are
\begin{eqnarray}
    P_{\mu e}^{\rm vac} &=& \sin^22\theta \sin^2\bigg(\dfrac{\Delta m^2 L}{4E}\bigg) \equiv  P_{ e\mu}^{\rm vac} \,,
    \nonumber\\
    P_{ee}^{\rm vac} &=& 1- P_{e\mu}^{\rm vac} \equiv P_{\mu\mu}^{\rm vac} \,,
\label{vacP}    
\end{eqnarray} 
where, $L$ is the distance traveled by neutrinos. For antineutrinos, $\xi\rightarrow-\xi$ and $\phi\rightarrow-\phi$ in Eq.~\ref{P}.
For standard oscillations and in absence of decay,  $P_{\mu e}^{\rm vac}$ = $P_{e \mu}^{\rm vac}$ and $P_{ee}^{\rm vac}$ = $P_{\mu \mu}^{\rm vac}$. This implies that $\cal CP$, $ \cal T$ and $\cal CPT$ are conserved in vacuum. However, in presence of decay, this no longer holds.
Eq.~\ref{P} implies that both $\cal CP$  is violated ($P_{\bar{e}\bar{\mu}} \neq P_{e\mu}$) and $\cal T$  is violated ($P_{\mu e} \neq P_{e\mu}$)  but   $\cal CPT$ is conserved \ie, $P_{\bar{\mu}\bar{e}} = P_{e\mu}$, $P_{\bar{e}\bar{e}} = P_{ee}$, $P_{\bar{\mu}\bar{\mu}} = P_{\mu\mu}$. There are two ways in which $\cal CP$ and $\cal T$ could be violated in presence of decay (with off-diagonal decay terms, $\eta \neq 0$) : 
\begin{itemize}
\item[(a)] 
$\phi \neq 0$  and  $\xi = 0$ (Majorana phase induced),  \item[(b)] $\xi \neq 0$ and $\phi = 0$ (decay induced). 
\end{itemize}

We note that the presence of off-diagonal term in the decay matrix plays a crucial role in revealing the dependence of the Majorana phase $\phi$ at the level of probabilities. 
By setting the off-diagonal terms in the decay matrix to zero, \ie, $\eta=0$ in Eq.~\ref{P},  we recover the  Dirac case (\ie, $\cal CP$ and $\cal T$ are conserved) as described in~\cite{Shafaq:2021lju}. 

In Fig.~\ref{fig:violations}, we plot the probability as a function of energy for a fixed baseline of $1300$ km. This baseline corresponds to an upcoming long baseline experiment based on Liquid Argon Time Projection Chamber (LArTPC) detector technology such as Deep Underground Neutrino Experiment (DUNE)~\cite{DUNE:2020ypp}. The top panel depicts the probability for the $\cal CP$ conjugate channels for contribution coming from off-diagonal decay term (right) or due to Majorana nature  of neutrinos (left). For comparison, the vacuum probability is also shown. 
The bottom panel shows the probability for $\cal T$ conjugate channels on the left and electron neutrino and muon neutrino survival probabilities on the right. It is evident from Fig.~\ref{fig:violations} that $\cal CP$ (and $\cal T$) could be violated either due to the off-diagonal term in the decay matrix or due to the Majorana phase. 
Further, the survival probabilities are flavour-dependent and not the same, as is expected from Eq.~\ref{P}.

\subsection{Leggett-Garg Inequalities and two flavour neutrino system}

Our intuition about the macroscopic world can be cast in terms of two principles which form the basis of  LGI~\cite{Leggett:1985zz} (for a review, see~\cite{Nori:2014lgi})
\begin{itemize}
\item[(a)] Macroscopic realism (MR) which implies that the measurement process reveals a well-defined pre-existing value. 
 \item[(b)] Non-invasive measurability (NIM) which states that we can measure this value without disturbing the system. 
 \end{itemize}
 These two assumptions are respected in the classical world. But, quantum mechanics  is based on superposition principle and  collapse of wave function under measurement which implies that these assumptions do not hold.

The problem of two flavour neutrino oscillations can be mapped onto a two-level quantum system in the ultra-relativistic approximation~\cite{Mehta:2009ea}. Once this mapping is clear, we can use our understanding of a general two-level quantum system which has been widely studied in the context  of LGI. For such a quantum system, the correlators are given by the symmetrized combination~\cite{Fritz_2010,Nori:2014lgi}
\begin{eqnarray}
    C_{ij} &=& \dfrac{1}{2}\langle \{\hat{Q}_{i}, \hat{Q}_j \}\rangle\, ,
\end{eqnarray}
{where $Q_i$ represents a dichotomic observable (\ie, takes values $+1$ or $-1$) defined over the Hilbert space of a given two-level quantum system.
For the simplest of LGI, three sets of runs are carried out experimentally to measure the two-time correlation function $C_{ij}$.} We can parameterize the qubit operators as $\hat{Q}_i = \vec{a_i} \cdot \vec{\sigma}$ where $\vec{\sigma}$ is the vector of Pauli matrices, $\vec{a_i}$ is a unit vector. Using the identity, 
${(\vec a_i \cdot \vec \sigma) ( \vec a_j \cdot \vec \sigma)} = \vec a_i \cdot \vec a_j + i(\vec a_i \times \vec a_j) \cdot  \vec \sigma $, 
and the fact that the vectors  $\vec a_i$ all lie in a plane, we obtain

\begin{eqnarray}
   \frac{1}{2}\langle \{\hat{Q}_{i},  \hat{Q}_j \}\rangle &=& \vec{a}_{i}\cdot \vec{a}_{j}\, .
\label{cija} 
\end{eqnarray}
For the $n^{th}$ order LGI parameter, we can write 
\begin{eqnarray}
K_n &=& \sum_{m=1}^{n-1} {C}_{m,m+1} - {C}_{1,n}\, . 
\label{Kn}
\end{eqnarray}
We can express this $K_n$ as 
\begin{eqnarray}
K_n &=& \sum_{m=1}^{n-1} \vec {a}_m \cdot\vec{a}_{m+1} - \vec{a}_1 \cdot\vec{a}_n\nonumber\\
&=& \sum_{m=1}^{n-1} \cos \theta_m - \left(\cos \sum_{m=1}^{n-1} \theta_m \right)\, ,
\label{Kn1}  
\end{eqnarray}

where $\theta_m$ is the angle between $\vec {a}_m$ and $\vec{a}_{m+1}$ and the angle between any pair of $\vec a_i$ and $\vec a_j$  are equal i.e., $\theta_{ij} \equiv \theta_m$. 
We can write generalised version of LGI as follows
\begin{eqnarray}
\label{classicalbound}
    -n\le K_n \le n-2 &\quad \quad  &   n\ge 3, \textrm{odd}   \nonumber\\
   -(n-2)\le K_n \le n-2 &\quad \quad &     n\ge 4,\textrm{even} 
\end{eqnarray} 

For a qubit, we can maximise $K_n$ by setting all angles, $\theta_m = \pi/n$ and obtain
\begin{eqnarray}
    K_{n}^{max} &=& n \cos \frac{\pi}{n}\, .
\end{eqnarray}

It  follows that 
\begin{eqnarray}
\label{upperbound}
    K_{3}^{max} = \frac{3}{2} \quad;\quad K_{4}^{max} = 2 \sqrt{2} \quad;\quad K_{5}^{max} = \frac{5}{4} (1+\sqrt{5}) \quad;\quad  K_{6}^{max} = 3 \sqrt{3}\, . 
\end{eqnarray}
These correspond to the Luder's bound or temporal Tsirelson bound~\cite{Cirelson:1980ry,Budroni2013BoundingTQ,budroni2014temporal}. $C_{ij}$ can be expressed in terms of  joint probabilities~\cite{Nori:2014lgi} 
\begin{eqnarray}\centering
    C_{ij} &=& \sum _{\hat Q_i \hat Q_j = \pm1} \hat Q_i \hat Q_j \mathbb{P}_{\hat Q_i \hat Q_j} (t_i,t_j)\, ,
\end{eqnarray}
where  $\mathbb{P}_{\hat Q_i \hat Q_j} (t_i,t_j)$ is the joint probability of obtaining the results $\hat Q_i$ and $\hat Q_j$ from successive measurements at times $t_i$ and $t_j$ respectively. Considering a muon neutrino $|\nu_\mu \rangle$ on which measurements are made at times $t_{i}$, the two flavour case, the $C_{12}$~\cite{Gangopadhyay:2013aha} can be written as
\begin{eqnarray}
    C_{12} &=& \mathbb{P}_{\nu_{e} \nu_{e}}(t_1,t_2)-\mathbb{P}_{\nu_{e} \nu_{\mu}}(t_1,t_2)-\mathbb{P}_{\nu_{\mu} \nu_{e}}(t_1,t_2)+\mathbb{P}_{\nu_{\mu} \nu_{\mu}}(t_1,t_2) \, , 
\label{jointP}    
\end{eqnarray}
where $\mathbb{P}_{\nu_{\alpha} \nu_{\beta}} (t_1,t_2)  
= P_{\mu \alpha} (t_1) P_{\alpha \beta} (\Delta t)$  is the joint probability of obtaining  neutrino in state $|\nu_\alpha\rangle$ at time $t_{1}$ and in state $|\nu_\beta \rangle $ at time $t_{2}$. This approach has been extended to three-flavour neutrino oscillation in Ref.~\cite{Gangopadhyay:2017nsn}.

\begin{figure}[ht!]
\includegraphics[width=1\textwidth]{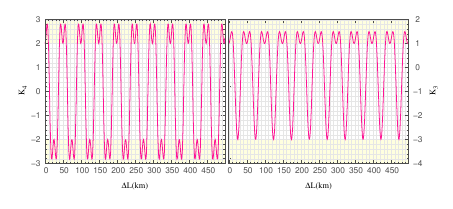} 
\caption{\label{k3k4nu} $K_3$ and $K_4$  plotted as a function of  $\Delta L$ for standard two flavour neutrino oscillations. The shaded regions represent quantum regime. 
Here, $\Delta m^2 = 2.5\times 10^{-3}$  eV$^2$, $\theta = 45^{\circ}$ and $E=50$ MeV.}
\end{figure}

The correlation function for the standard two flavour neutrino oscillations is (for details, see~\cite{Formaggio:2016cuh, Shafaq:2020sqo, Shafaq:2021lju})
\begin{eqnarray}
    {C}_{ij} &=& 1-2\sin^2{2\theta} \sin^2{\psi_{ij}}\, , 
\end{eqnarray}

where $\theta$ is the mixing angle in vacuum and $\psi_{ij}$ is given by
\begin{eqnarray}
    \psi_{ij} &=&  \frac{\Delta m^2}{4E} (t_j - t_i)\, .
\label{psi}
\end{eqnarray}

Assuming equal time intervals, {\ie}, $t_{m+1} -t_m = \tau$ (which corresponds to the stationarity condition), we obtain
\begin{eqnarray}
\label{2levelk3k4}
K_3 &=& 1 - 2 \sin^2 2\theta \left[ 2 \sin ^2 \dfrac{\Delta m^2 \tau}{4E}   -  \sin ^2 \dfrac{2\Delta m^2 \tau}{4E}  \right]\, ,\nonumber\\
K_4 &=& 2 - 2 \sin^2 2\theta \left[ 3 \sin ^2 \dfrac{\Delta m^2 \tau}{4E}   -  \sin ^2 \dfrac{3\Delta m^2 \tau}{4E}   \right]\, .
\end{eqnarray}
In the ultra-relativistic limit, we can replace $\tau$ by $\Delta L$ = $L_i$ $-$ $L_j$, where $L_i$ and $L_j$ are the fixed distances from the neutrino source where the measurements occur.

In Fig.~\ref{k3k4nu}, We  depict $K_3$ and $K_4$  as a function of $\Delta L$ for  standard two flavour neutrino oscillations. It can be noted that $K_3$ and $K_4$ (see Eq.~\ref{2levelk3k4}) exceed their respective classical bounds (Eq.~\ref{classicalbound}) however they respect the maximum upper bounds (Eq.~\ref{upperbound}).
Next, we will study the impact of neutrino decay on two flavour neutrino oscillations and its implications for LGI.

\section{Non-Hermitian neutrino Hamiltonian and  implications for LGI - role of the Majorana phase} \label{sec:3}
\begin{figure}[htb!]
\centering
\includegraphics[width=3.7in]{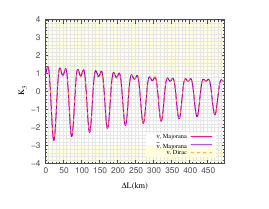} 
\hskip -1.2in
\includegraphics[width=3.7in]{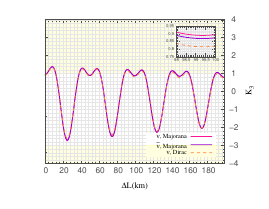}
\caption{\label{k3} $K_3$ plotted as function of $\Delta L$  for the case of  Dirac and Majorana neutrinos.  $K_3$ depends on neutrinos or anti-neutrinos for the Majorana case (Eq.~\ref{C_neutrino}). The left and right plots are for two different ranges of $\Delta L$ to show the effects clearly. 
In the inset  (shown in the right panel), $K_3$ is plotted as function of $\Delta L$ near the location of the peak at around $\Delta L = 100$ km. 
The parameter values are taken to be the same as those given in the caption of Fig.~\ref{fig:violations}.}
\end{figure}
\begin{figure}[ht!]
\centering
\includegraphics[width=3.7in]{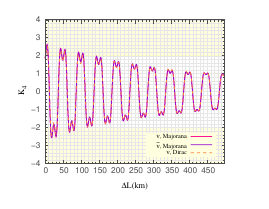} 
\hskip -1.2in
\includegraphics[width=3.7in]{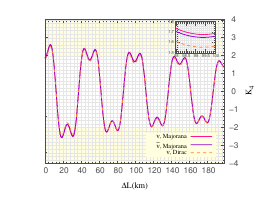}
\caption{\label{k4} Same as Fig.~\ref{k3} but for $K_4$.}
\end{figure}

Using the prescription to compute LGI parameter (Eq.~\ref{jointP}) for  the case of $\cal PT$ symmetric non-Hermitian Hamiltonian given by Eq.~\ref{P}, we get the following
\begin{figure}[ht!]
\centering
\includegraphics[width=5.5in]{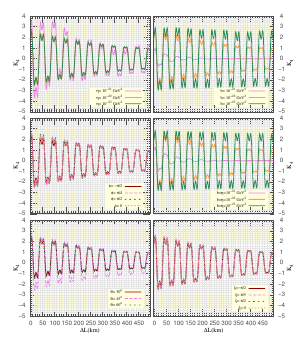}
\caption{\label{variations} Parameter dependence on $K_4$ is shown  as a function of $\Delta L$. 
Here, $\Delta L$ varies between $0-500$ km and  $E=50$ MeV. The parameter values are taken to be the same as those given in the caption of Fig.~\ref{fig:violations}.}
\end{figure}
\begin{eqnarray}
    C_{12} &=& e^{-2b(L+\Delta L)}\Big[1+x\left[\mathcal{A}(L)+\mathcal{A}(\Delta L)\right]-2x\left[\mathcal{A}
    (L)P_{e\mu}^{\rm vac}(\Delta L)+ \mathcal{A}(\Delta L)P_{e\mu}^{\rm vac}(L)\right]
    \nonumber\\
    &&  \mp  2y[\mathcal{B}(L)-\mathcal{B}(\Delta L)]-2 P_{e\mu}^{\rm vac}(\Delta L)\Big] \,,\nonumber\\    
    C_{23}&=&e^{-2b(L+2\Delta L)}\Big[1+x[\mathcal{A}(\Delta L)+\mathcal{A}(L+\Delta L)]-2x[\mathcal{A}(\Delta L)P_{e\mu}^{\rm vac}(L+\Delta L)\nonumber\\
    &&  + \mathcal{A}(L+ \Delta L)P_{e\mu}^{\rm vac}(\Delta L)] \mp 2y[\mathcal{B}(L+ \Delta L)-\mathcal{B}(\Delta L)]-2 P_{e\mu}^{\rm vac}(\Delta L)\Big] \,, \nonumber\\
    C_{34} &=& e^{-2b(L+3\Delta L)}\Big[1+x[\mathcal{A}(\Delta L)+\mathcal{A}(L+2\Delta L)]-2x[\mathcal{A}(\Delta L)P_{e\mu}^{\rm vac}(L+2\Delta L)\nonumber\\
    && + \mathcal{A}(L+ 2\Delta L)P_{e\mu}^{\rm vac}(\Delta L)] \mp 2y[\mathcal{B}(L+ 2\Delta L)-\mathcal{B}(\Delta L)] -2 P_{e\mu}^{\rm vac}(\Delta L)\Big] \,,\nonumber\\
    C_{13} &=&e^{-2b(L+2\Delta L)}\Big[1+x[\mathcal{A}(2\Delta L)+\mathcal{A}(L)]-2x[\mathcal{A}(2\Delta L)P_{e\mu}^{\rm vac}(L)+ \mathcal{A}(L)P_{e\mu}^{\rm vac}(2\Delta L)]\nonumber\\
    && \mp  2y[\mathcal{B}(L)-\mathcal{B}(2\Delta L)]-2 P_{e\mu}^{\rm vac}(2\Delta L)\Big] \,, \nonumber\\ 
    C_{14}&=&e^{-2b(L+3\Delta L)}\Big[1+x[\mathcal{A}(3\Delta L)+\mathcal{A}(L)]-2x[\mathcal{A}(3\Delta L)P_{e\mu}^{\rm vac}(L)+ \mathcal{A}(L)P_{e\mu}^{\rm vac}(3\Delta L)]\nonumber\\
    &&  \mp \, 2y[\mathcal{B}(L)-\mathcal{B}(3\Delta L)]-2 P_{e\mu}^{\rm vac}(3\Delta L)\Big] \,.
\label{C_neutrino}
\end{eqnarray}

\begin{figure}[ht!]
\centering
\includegraphics[width=5.5in]{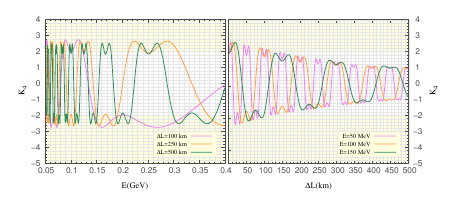}
\caption{\label{variations2} The dependence of $K_4$ on $\Delta L$ (left) and $E$ (right) is depicted in this figure. 
The parameter values are taken to be the same as those given in the caption of Fig.~\ref{fig:violations}.}
\end{figure}
Here, $x=\eta \cos(\xi-\phi)$ , $y=\eta \sin(\xi-\phi)$, values of $\mathcal{A}$ and $\mathcal{B}$ are given by Eq.~\ref{AB} and value of $P_{e\mu}^{\rm vac}$ is given in Eq.~\ref{vacP}. In the above equations, the upper  sign ($-$) is for neutrinos and lower sign ($+$) is for anti-neutrinos. Since $\eta$ is very small, we neglect terms of $\mathcal{O}(\eta^2)$ to make our expressions more compact~\cite{Dixit:2022izn}. The corresponding expressions for $K_3$ and $K_4$   can be computed using the prescription given in Eq.~\ref{Kn}. In order to obtain the corresponding expression for Dirac neutrinos, we put $\eta=0$ in Eq.~\ref{C_neutrino}. We note that we recover the expressions for $K_3$ and $K_4$ given in Ref.~\cite{Shafaq:2021lju} in this case.  

In Fig.~\ref{k3}, $K_3$ plotted as function of $\Delta L$  for the case of  Dirac and Majorana neutrinos. As can be seen, $K_3$ depends on neutrinos or anti-neutrinos for the Majorana case (Eq.~\ref{C_neutrino}). The left and right plots are for two different ranges of $\Delta L$ (the right plot is a zoomed in version of the left plot) to demonstrate the effects clearly. 
In the inset  (shown in the right panel), $K_3$ is plotted as function of $\Delta L$ near the location of the peak at around $\Delta L = 100$ km. 

Fig.~\ref{k4} depicts the behaviour of the LGI parameter $K_4$ plotted as function of $\Delta L$. The dependence of $K_4$ on different parameters is shown in Fig.~\ref{variations} and Fig.~\ref{variations2}. In Fig.~\ref{variations},  $K_4$ is plotted as function of $\Delta L$ for different values of decay parameters  ($b$, $\eta$, $\xi$), Majorana Phase ($\phi$) and the mixing angle ($\theta$). In Fig.~\ref{variations2}, 
 $K_4$ is plotted as a function of energy for different values of $\Delta L$ (left) and  $\Delta L$ for different values of $E$ (right).   From Fig.~\ref{variations} and Fig.~\ref{variations2}, we can deduce the following parameters that maximize $K_4$. These are:  $\theta = 45^{\circ}$, $b=10^{-22}$ GeV$^2$,  $\eta\ = 10^{-22}$ GeV$^2$, $\xi\ = \pi/5$, $\phi\ = \pi/4$ and $E=50$ MeV.

\begin{figure}[ht!]
    \includegraphics[width=1\textwidth]{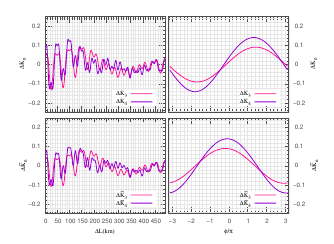}
    \caption{\label{deltaK} $\Delta K_3$ and $\Delta K_4$ plotted as function of $\Delta L$ (left) and $\phi$ (right). 
    The parameter values are taken to be the same as those given in the caption of Fig.~\ref{fig:violations} and we have taken $\Delta L=100$ km in the right panel.
   }
\end{figure}

In order to quantify the distinction between Dirac and Majorana case, we  define
\begin{eqnarray}
    \Delta K_n &=& K_{n}^{\nu,M} - K_{n}^{\nu,D}\, , \nonumber\\
    \Delta\bar{K_n} &=& K_{n}^{\nu,M} - \bar K_{n}^{\nu,D}\, ,
    \label{deltaKf}
\end{eqnarray}
where $K_{n}^{\nu,M}$ ($\bar{K_n}^{\nu,M}$) corresponds to Majorana neutrino case  (Majorana  anti-neutrino case) and $K_{n}^{\nu,D}$  corresponds to the Dirac case. We plot  $\Delta K_3$ and $\Delta K_4$ as a function of $\Delta L $ (left panel) and the Majorana phase $\phi$ (right panel)  in Fig~\ref{deltaK}. The dependence on the Majorana phase $\phi$ is guided by oscillatory terms involving  $(\xi-\phi)$ and in general, we do not expect symmetric behaviour about $\phi=0$.

\begin{figure}[ht!]
    \includegraphics[width=1\textwidth]{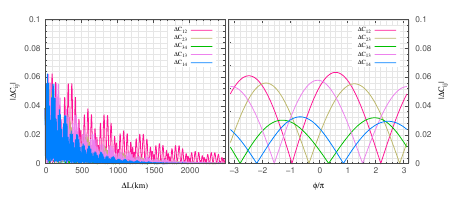} 
    \caption{\label{deltaCij} $\Delta C_{ij}$ plotted as function of $\Delta L$ (left) and Majorana phase, $\phi$ (right).  Similar pattern of $\Delta C_{ij}$ is seen for anti-neutrinos. }
\end{figure}

Interestingly, we find that $|\Delta K_3| \sim 10 \%$ and $|\Delta K_4| \sim 15\%$ thereby implying that $K_4$ allows for a better distinction between Dirac and Majorana cases   for favourable choice of parameters.   This particular trend is generally true for anti-neutrinos as well. To understand why this is the case, we plot   individual contributions of $\Delta C_{ij}$s in Fig.~\ref{deltaCij}. It can be seen that 
(i)  there is a large contribution coming from $C_{23}$ which appears in $K_4$ with a plus sign, and (ii) the contribution of 
$C_{13}$ is larger than $C_{14}$. As both appear with a minus sign in $K_3$ and $K_4$ respectively, it naturally makes $K_3$ smaller than $K_4$. These two effects collectively give $\Delta K_4$ an advantage over $\Delta K_3$ when it is used for distinguishing Dirac and Majorana cases. 

\section{Conclusion}\label{Sec:4}

Study of temporal correlations in the form of LGI has attracted significant attention in recent times both in the context of two and three flavour neutrino oscillations~\cite{Gangopadhyay:2013aha,Gangopadhyay:2017nsn,Naikoo:2017fos,Naikoo:2019eec,Naikoo:2019gme,Formaggio:2016cuh,Fu:2017hky,Shafaq:2020sqo,Sarkar:2020vob,Yadav:2022grk}.  It should be noted that while different dichotomic observables have been employed in these studies, the  dynamics was restricted to be Hermitian in  these studies.

It is believed that Majorana phase appearing in the two flavour neutrino mixing matrix can not have any effect on the neutrino oscillation probabilities~\cite{Giunti:2010ec}. It should be noted that this holds as long as the dynamics is Hermitian. However, if we replace the condition of self-adjointness by enforcing $\cal PT$ symmetry, then the above claim does not hold. One possible way to realize non-Hermitian $\cal PT$ symmetric Hamiltonian is to incorporate neutrino decay along with neutrino oscillation. 
In order that the Majorana phase appears in two flavour oscillation probability, it is essential that off-diagonal terms in the decay matrix are non-zero~\cite{Dixit:2022izn}. For the scenario of neutrino oscillation and decay in vacuum  (with off-diagonal terms in the decay matrix), we explore the violation LGI and show that it is possible to discriminate between Dirac and Majorana neutrinos. If the decay matrix is diagonal as considered in~\cite{Shafaq:2021lju}, the Majorana phase ceases to play a role at the level of probability, as expected. 

Presently, we do not know if neutrinos are Dirac or Majorana particles~\cite{Majorana2020} and in future neutrinoless double $\beta$-decay process is expected to provide conclusive evidence. Moreover, exploring effects of CP violation due to Majorana phase is quite challenging~\cite{deGouvea:2002gf}. 
Complementing the results of neutrinoless double $\beta$-decay experiments, several interesting proposals have emerged that could potentially allow us to probe the nature of neutrinos as well as CP violating effects due to Majorana phases using cosmological probes~\cite{Hernandez-Molinero:2022zoo} or  ideas of geometric phases and quantum decoherence~\cite{Dajka:2011zz,Capolupo:2018hrp,Buoninfante:2020iyr,Capolupo:2020hqm}.  
Richter et al.~\cite{Richter:2017toa} considered quantum decoherence in the density matrix formalism along with matter effects and proposed  that LGI could be used as a probe of the nature of neutrinos i.e., whether they are of Dirac-type or Majorana-type. The claim was based on the assumption that matter effects played a role and  the off-diagonal terms in the decoherence matrix were non-zero.  
Recently, 
King et. al~\cite{King:2023cgv} have used the spectrum of gravitational waves to probe the nature of neutrinos. 

With the goal of probing the nature of neutrinos, in the present work, we consider a scenario in which we exploit the non-Hermiticity of the Hamiltonian (by invoking decay with neutrino oscillations) and show that even in vacuum,  with non-zero off-diagonal terms in the decay matrix, it is possible to distinguish Dirac and Majorana cases by studying the extent of violation of LGI. We quantify this in terms of  $\Delta K_3$ and $\Delta K_4$ (Eq.~\ref{deltaKf}) and  find that  $|\Delta K_3| \sim 10\%$ and $|\Delta K_4| \sim 15\%$ which means that $K_4$ allows for a better discrimination between Dirac and Majorana case for favourable choice of parameters.

\section*{Acknowledgements}
BS would like to thank PM for the warm hospitality during her visits to JNU, New Delhi. BS acknowledges Sabila Parveen for  discussions and help with Gnuplot software. PM would like to thank Dibya Chattopadhyay and Amol Dighe for discussions. The work of PM is partially supported by the European Union’s Horizon 2020 research and innovation programme under the Marie Skodowska-Curie grant agreement No 690575 and 674896. 
%

\cleardoublepage

\bibliographystyle{ieeetr} 
{\footnotesize
\bibliography{biblography}}

\end{document}